\documentclass{llncs}
\usepackage{makeidx}
\usepackage[pdftex]{graphicx, color}
\usepackage{amsmath,amssymb}
\usepackage{enumerate}
\usepackage[colorlinks=true, allcolors=blue]{hyperref}

\newcommand{\E}{E \,}
\newcommand{\PP}{P}
\newcommand{\dto}{\xrightarrow{d}}
\newcommand{\pto}{\xrightarrow{\PP}}

\newcommand{\trccThe}{c_{\rm tr}} 
\newcommand{\diccThe}{c_{\rm di}} 
\newcommand{\diccEmp}{{\cal C}_{\rm di}} 

\begin{document}

\mainmatter
\title{Diclique clustering in a directed random graph}
\titlerunning{Diclique clustering}
\author{Mindaugas Bloznelis\inst{1} \and Lasse Leskel\"a\inst{2}
}
\authorrunning{M. Bloznelis and L. Leskel\"a}
\tocauthor{Mindaugas Bloznelis, Lasse Leskel\"a}
\institute{
Vilnius University, Lithuania, \ \href{http://www.mif.vu.lt/~bloznelis/}{\nolinkurl{www.mif.vu.lt/~bloznelis/}}
\and 
Aalto University, Finland,  \ \href{http://math.aalto.fi/\~lleskela/}{\nolinkurl{math.aalto.fi/\~lleskela/}}
}
\maketitle

\begin{abstract}
We discuss a notion of clustering for directed graphs, which describes how likely two followers of a  node are to follow a 
common target.  The associated network motifs, called dicliques or bi-fans, have been found to be key structural components in various real-world networks.  We introduce a two-mode statistical network model consisting of actors and auxiliary attributes, 
where an actor $i$ decides to follow an actor $j$ whenever $i$ demands an attribute supplied by $j$. 
We show that the digraph admits nontrivial clustering properties of the aforementioned type, 
as well as power-law indegree and outdegree distributions. 
\end{abstract}

\keywords{intersection graph, two-mode network, affiliation network, digraph, diclique, bi-fan, complex network}

\section{Introduction}

\subsection{Clustering in directed networks}

Many real networks display a tendency to cluster, that is, 
to form dense local neighborhoods in a globally sparse graph. 
In an undirected social network this may be phrased as: 
\emph{your friends are likely to be friends.}  
This feature is typically quantified in terms 
of local and global clustering coefficients measuring how likely two neighbors of a node are neighbors \cite{Newman_2003_Structure,Scott_2012,Szabo_Alava_Kertesz_2004,Watts_Strogatz_1998}. In directed networks there are many ways to define the concept of clustering, for example by considering the thirteen different ways that a set of three nodes may form a weakly connected directed graph \cite{Fagiolo_2007}.

\begin{figure}[h]
\begin{center}
\includegraphics[width=40mm]{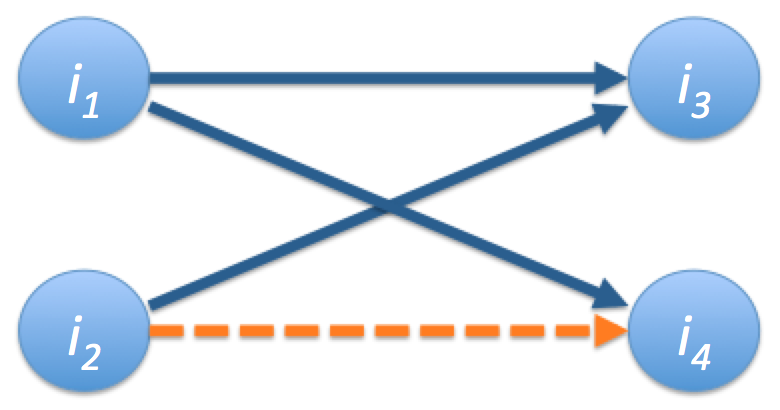}
\end{center}
\caption{\label{fig:Diclique} Forming a diclique by adding a link $i_2 \to i_4$.}
\end{figure}

In this paper we discuss 
a new type of clustering concept which is motivated by directed online 
social networks, where a directed link 
$i \to j$ means that an actor $i$ 
follows actor $j$. 
In such networks a natural way to describe clustering is to say that 
{\it your followers are likely to follow common targets}. 
When the topology of the network is unknown and 
modeled as a random graph distributed according to a 
probability measure $\PP$, 
the above statement can be expressed as
\begin{equation}
 \label{eq:DicliqueClustering}
 \PP( i_2 \to i_4 \, \bigr|\, i_1 \to i_2, \, i_1 \to i_3, \, i_2 \to i_3)
 \ > \
 \PP( i_2 \to i_4 ),
\end{equation}
where 'you' corresponds to actor $i_3$. Interestingly, 
the conditional probability on the 
left can  stay bounded away from zero even for sparse random digraphs~\cite{Leskela_2015_Istanbul}. The associated subgraph (Fig.~\ref{fig:Diclique}) is called a \emph{diclique}. 
Earlier experimental studies have observed that dicliques (a.k.a.\ bi-fans) 
constitute a key structural motif in gene 
regulation networks 
\cite{Milo_Shen-Orr_Itzkovitz_Kashtan_Chklovskii_Alon_2002}, 
citation networks, and several types of online social networks 
\cite{Zhang_Lu_Wang_Zhu_Zhou_2013}. 

Motivated by the above discussion, we define a global \emph{diclique clustering coefficient} of a finite directed graph $D$ with an adjacency matrix $(D_{ij})$ by
\begin{equation}
 \label{eq:ClusteringAdj}
 \diccEmp(D)
 =
 \frac
{\sum_{(i_1,i_2,i_3,i_4)} D_{i_1,i_3} D_{i_1,i_4} D_{i_2,i_3} D_{i_2,i_4}}
{\sum_{(i_1,i_2,i_3,i_4)} D_{i_1,i_3} D_{i_1,i_4} D_{i_2,i_3}},
\end{equation}
where the sums are computed over all ordered quadruples of distinct nodes. 
It provides
an empirical 
counterpart to the conditional probability (\ref{eq:DicliqueClustering}) 
in the sense 
that 
 the ratio in (\ref{eq:ClusteringAdj}) defines the conditional 
probability
\begin{equation}
 \label{eq:ClusteringCond}
 \PP_D\bigl( I_2 \to I_4 \, \, \bigl| \, \, I_1 \to I_3, \, I_1 \to I_4, \, I_2 \to I_3\bigr),
\end{equation}
where  $\PP_D$ refers to the distribution of the 
random quadruple $(I_1,I_2,I_3,I_4)$ 
sampled uniformly at random among all ordered quadruples of distinct nodes~in~$D$.

To quantify diclique clustering among the followers of a selected actor $i$,
we may define a local diclique clustering coefficient by
\begin{equation}
 \label{eq:ClusteringAdj+}
 \diccEmp(D,i)
 =
 \frac
 {\sum_{(i_1,i_2,i_4)} D_{i_1,i} D_{i_1,i_4} D_{i_2,i} D_{i_2,i_4}}
 {\sum_{(i_1,i_2,i_4)} D_{i_1,i} D_{i_1,i_4} D_{i_2,i}},
\end{equation} 
where the sums are computed over all ordered triples of distinct nodes excluding $i$. We remark that $\diccEmp(D,i) = \PP_D\bigl( I_2 \to I_4 \, \bigl| \, I_1 \to I_3, I_1 \to I_4, I_2 \to I_3,\, I_3=i\bigr)$.

\begin{remark}
By replacing $\to$ by $\leftrightarrow$  in \eqref{eq:ClusteringCond}, we see that the
analogue of the above notion for undirected graphs corresponds to predicting how likely the endpoints
of the 3-path $I_2 \leftrightarrow I_3 \leftrightarrow I_1 \leftrightarrow I_4$ are linked together.
\end{remark}

\subsection{A directed random graph model}
\label{sec:Model}

Our goal is to define a parsimonious yet powerful statistical model of a directed social network which displays diclique clustering properties as discussed in the previous section. Clustering properties in many social networks, such as
movie actor networks or scientific collaboration networks, are explained by underlying bipartite structures relating actors to movies and scientists to papers \cite{Guillaume_Latapy_2004,Newman_Strogatz_Watts_2001}. Such networks are naturally modeled using directed or undirected random intersection graphs \cite{Bloznelis_2010_Random,Bloznelis_Godehardt_Jaworski_Kurauskas_Rybarczyk_2015,Deijfen_Kets_2009,Frieze_Karonski_2016,Karonski_Scheinerman_Singer-Cohen_1999}.

\begin{figure}[h]
\begin{center}
\includegraphics[width=60mm]{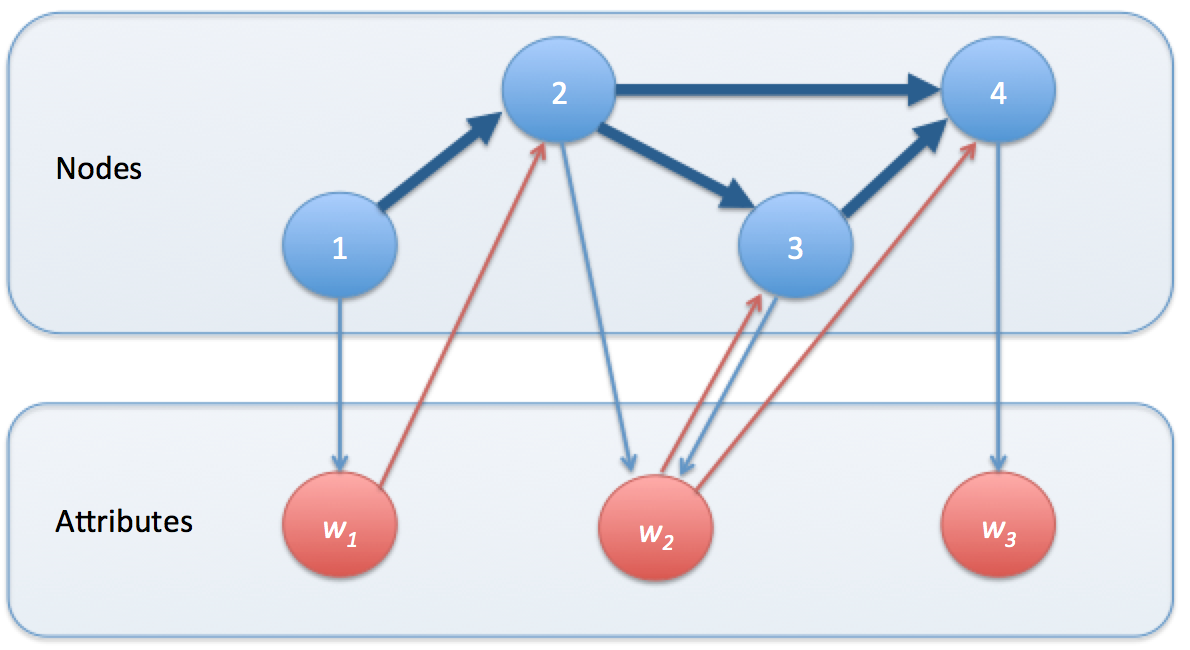}
\end{center}
\caption{\label{fig:RIG} Node 1 follows node 2, because 1 demands attribute $w_1$ supplied by 2.}
\end{figure}

A directed intersection graph on a node set $V=\{1,\dots, n\}$ is constructed with the help of an auxiliary set of attributes $W = \{w_1,\dots,w_m\}$ and a directed bipartite graph $H$ with bipartition $V\cup W$, which models how nodes (or actors) relate to attributes. We say that actor $i$ \emph{demands} (or follows) attribute $w_k$ when $i \to w_k$, and \emph{supplies} it when $i \leftarrow w_k$. The directed intersection graph $D$ induced by $H$ is the directed graph on $V$ such that $i \to j$ if and only if $H$ contains a path $i\to w_k \to j$, or equivalently, $i$ demands one or more attributes supplied by $j$ (see Fig.~\ref{fig:RIG}). For example, in a citation network the fact that an author $i$ cites a paper $w_k$ coauthored by $j$, corresponds to $i \to w_k \to j$.

We consider a random bipartite digraph $H$ where the pairs $(i,w_k)$, $i\in V$, $w_k\in W$ establish adjacency relations independently of each other.  That is, the bivariate binary random vectors $({\mathbb I}_{i\to k},\, {\bf I}_{k\to i})$, $1\le i\le n$, $1\le k\le m$, are { stochastically independent}. Here ${\mathbb  I}_{i\to k}$ and ${\bf I}_{k\to i}$ stand for the indicators of the events that links $i\to w_k$ and $w_k\to i$ are present in $H$. We assume that every pair  $(i,w_k)$ is assigned a triple of probabilities
\begin{equation}\label{pikqik}
 p_{ik}=\PP(i\to w_k),
\quad
\
q_{ik}=\PP(w_k\to i),
\quad
\
r_{ik}=\PP(i\to w_k,\, w_k\to i).
\end{equation}
Note that, by definition, $r_{ik}$ satisfies the inequalities
\begin{equation}\label{rrr+}
 \max\{p_{ik}+q_{ik}-1,0\}
 \ \le \ r_{ik}
 \ \le \ \min\{p_{ik}, q_{ik}\}.
\end{equation}
A collection of triples 
$\{(p_{ik}, \, q_{ik}, \, r_{ik}),\,1\le i\le n,\, 1\le k\le m\}$
defines the distribution of a random bipartite digraph $H$.

We will focus on a fitness model where every node $i$ is prescribed 
a pair of weights $x_i,y_i\ge 0$ 
modelling the demand and supply intensities of  $i$. 
Similarly, every attribute $w_k$ is prescribed  a weight $z_k>0$ 
modelling its relative popularity. 
Letting 
\begin{equation}\label{pq}
p_{ik}=\min\Bigl\{1, \gamma x_i z_k\Bigr\}
\qquad
{\text{and}}
\qquad
q_{ik}=\min\Bigl\{1,\gamma y_i z_k\Bigr\},
\qquad
i,k\ge 1,
\end{equation}
we obtain link probabilities proportional to respective weights. 
Furthermore, we assume that 
\begin{equation}\label{rrr}
 r_{ik}=r(x_i,y_i,z_k,\gamma), 
\qquad
i,k\ge 1,
\end{equation}
for some function $r\ge 0$ satisfying (\ref{rrr+}). Here $\gamma>0$ is a parameter, defining the link density in $H$, which generally depends on $m$ and $n$. Note that $r$ defines the correlation between reciprocal links $i\to w_k$ and $w_k\to i$. 
For example, by letting $r(x,y,z,\gamma) = (\gamma x z \wedge 1)(\gamma y z \wedge 1)$, we obtain a random bipartite digraph with independent links.

We will consider weight sequences having desired statistical properties for complex network modelling. For this purpose we assume that the node and attributes weights are realizations of random sequences $X=(X_i)_{i \ge 1}$, $Y=(Y_i)_{i \ge 1}$, and $Z=(Z_k)_{k \ge 1}$, such that the sequences $\{(X_i,Y_i),\, i\ge 1\}$ and $\{Z_k,\, k\ge 1\}$ are mutually independent and consist of independent and identically distributed terms. The resulting random bipartite digraph is denoted by ${\cal H}$, and the resulting random intersection digraph by ${\cal D}$. We remark  that ${\cal D}$ extends the random intersection digraph model introduced in \cite{Bloznelis_2010_Random}.

\subsection{Degree distributions}
  
When $\gamma =(mn)^{-1/2}$ and $m,n \to \infty$, the random digraph ${\cal D}$ defined in Sec.~\ref{sec:Model} becomes sparse, having the number of links proportional to the number of nodes. Theorem \ref{T1} below describes the class of limiting distributions of the outdegree of a typical vertex $i$. We remark that for each $n$ the outdegrees $d_{+}(1)$,\dots, $d_{+}(n)$ are identically distributed.

To state the theorem, we let $\Lambda_1,\Lambda_2,\Lambda_3$ be mixed-Poisson random variables distributed according to
\[
 \PP(\Lambda_i = r)
 \ = \ \E \,e^{-\lambda_i}\frac{\lambda_i^r}{r!}, \quad r \ge 0,
\] 
where $\lambda_1=X_1\beta^{1/2}\E Z_1$, $\lambda_2=Z_1\beta^{-1/2}\E Y_1$, and $\lambda_3=X_1(\E Y_1)(\E Z_1^2)$. We also denote by $\Lambda_i^*$ a downshifted size-biased version of $\Lambda_i$, distributed according to
\[
 \PP(\Lambda_i^* = r) \ = \ \frac{r+1}{\E\Lambda_i}\PP(\Lambda_i=r+1), \quad r \ge 0.
\]
Below $\dto$ refers to convergence in distribution.

\begin{theorem}
\label{T1}
Consider a model with $n,m \to \infty$ and $\gamma=(nm)^{-1/2}$, and assume that $\E Y_1, \E Z_1^2  < \infty$.
\begin{enumerate}[(i)]
\item \label{ite:T11} If $m/n\to 0$ then $d_+(1) \dto 0$.
\item \label{ite:T12} If $m/n\to \beta$ for some $\beta\in (0,\infty)$, then
$
 d_+(1) \dto \sum_{j=1}^{\Lambda_1} \Lambda_{2,j}^*,
$
where $\Lambda_{2,1}^*, \Lambda_{2,2}^*, \dots$ are independent copies of $\Lambda_{2}^*$, and independent of $\Lambda_1$.

\item \label{ite:T13} If $m/n \to \infty$, then  $d_{+}(1) \dto \Lambda_3$.
\end{enumerate}
\end{theorem}

\begin{remark}
\label{rem:Indegree}
By symmetry, the results of Theorem~\ref{T1} extend to the indegree $d_{-}(1)$ when we redefine $\lambda_1=Y_1\beta^{1/2}\E Z_1$, $\lambda_2=Z_1\beta^{-1/2}\E X_1$, and $\lambda_3=Y_1(\E X_1)(\E Z_1^2)$.
\end{remark}

\begin{remark}
The limiting distributions appearing in Theorem~\ref{T1}:\eqref{ite:T12}--\eqref{ite:T13} admit heavy tails. This random digraph model is rich  enough to model power-law indegree and outdegree distributions, or power-law indegree  
and light-tailed outdegree distributions.
\end{remark}

\begin{remark}
The moment conditions in Theorem~\ref{T1} are not the sharpest possible. For example, in~\eqref{ite:T11} it is sufficient to assume that $\E Z_1 < \infty$.
\end{remark}

We note that  a related result for simple (undirected) random intersection graph has been shown in \cite{Bloznelis_Damarackas_2013}. Theorem \ref{T1} extends the result of \cite{Bloznelis_Damarackas_2013}
to digraphs.

\subsection{Diclique clustering}

We investigate clustering in the random digraph ${\cal D}$ defined in Sec.~\ref{sec:Model} by approximating the
(random) diclique clustering coefficient $\diccEmp({\cal D})$ defined in (\ref{eq:ClusteringAdj}) by a related nonrandom quantity
\begin{displaymath}
 \diccThe
:=
\PP(I_2\to I_4\, \bigr|\,  I_1 \to I_3, \, I_1 \to I_4, \, I_2 \to I_3 \bigr),
\end{displaymath}
where $(I_1,I_2,I_3,I_4)$ is a random ordered quadruple of distinct nodes chosen uniformly at random.
Note that here $\PP$ refers to two independent sources of randomness: 
the random digraph generation mechanism and the sampling of the nodes. Because the distribution of ${\cal D}$ is invariant with respect to a relabeling of the nodes, the above quantity can also be written as
\begin{displaymath}
 \diccThe \ = \ \PP\bigl( 2 \to 4 \, \bigr|\, 1 \to 3, \, 1 \to 4, \, 2 \to 3 \bigr).
\end{displaymath}
 We believe that under mild regularity conditions 
$\diccEmp({\cal D}) \approx \diccThe$, provided that $m$ and $n$ are sufficiently large. 
Proving this is left for future work.

Theorem~\ref{TDC} below shows that the random digraph ${\cal D}$ admits a {\it nonvanishing} clustering coefficient 
$\diccThe$ when the intensity $\gamma$ is inversely proportional to the number of attributes. For example, by choosing $\gamma =(nm)^{-1/2}$ and letting $m,n\to \infty$ so that $m/n\to \beta>0$, we obtain a sparse random digraph with tunable clustering coefficient $\diccThe$ and limiting degree distributions defined by Theorem~\ref{T1} and Remark~\ref{rem:Indegree}.

\begin{theorem}
\label{TDC}
Assume that $m \to \infty$ and $\gamma m \to \alpha$ for some constant $\alpha \in (0,\infty)$, and that
$\E X_1^3, \E Y_1^3, \E Z_1^4<\infty$. Then
\begin{equation}
 \label{DC}
 \diccThe
\to
 \left( 
1
+ 
\alpha 
\left( \frac{\E X_1^2}{\E X_1}+\frac{\E Y_1^2}{\E Y_1}\right)
\frac{(\E Z_1^2)(\E Z_1^3)}{\E Z_1^4}
+
\alpha^2
\frac{\E X_1^2}{\E X_1}
\frac{\E Y_1^2}{\E Y_1}\frac{(\E Z_1^2)^3}{\E Z_1^4}
\right)^{-1}.
\end{equation}
\end{theorem}

\begin{remark}
\label{rem:Five} 
When $\E X_1^4, \E Y_1^4, \E Z_1^4<\infty$, the argument in the proof of Theorem~\ref{TDC}
allows to conclude that $\diccThe \to 0$ when $\gamma m\to \infty$.
\end{remark}

To investigate clustering among the followers of a particular ego node~$i$, we study a theoretical analogue of the local diclique clustering coefficient $\diccEmp(D,i)$ defined in (\ref{eq:ClusteringAdj+}). By symmetry, we may relabel the nodes so that $i=3$. We will consider the weights of node 3 as known and analyze the conditional probability
\[
 \diccThe(X_3,Y_3)
 \ = \ \PP_{X_3,Y_3}(2\to 4 \, \, \bigr|\, \,  1 \to 3, \, 1 \to 4, \, 2 \to 3 \bigr),
\]
where $\PP_{X_3,Y_3}$ refers to the conditional probability given $(X_3,Y_3)$. Actually, we may we may replace $\PP_{X_3,Y_3}$ by $\PP_{Y_3}$ above, because all events appearing on the right are independent of $X_3$.

One may also be interested in analyzing the conditional probability
\[
 \diccThe(X,Y)
 \ = \ \PP_{X,Y}(2\to 4 \, \, \bigr|\, \,  1 \to 3, \, 1 \to 4, \, 2 \to 3 \bigr),
\]
where $\PP_{X,Y}$ refers to the conditional probability given the values of all node weights $X=(X_i)$ and $Y=(Y_i)$. Again, we may replace $\PP_{X,Y}$ by $\PP_{X_1,X_2,Y_3,Y_4}$ above, because the events on the right are independent of the other nodes' weights. More interestingly, $\diccThe(X,Y)$ turns out to be asymptotically independent of $X_2$ and $Y_4$ as well in the sparse regime.

\begin{theorem}
\label{TLDC}
Assume that $m \to \infty$ and $\gamma m \to \alpha$ for some constant $\alpha \in (0,\infty)$.
\begin{enumerate}[(i)]
\item \label{ite:TLDC1} If $\E X_1^3, \E Y_1^3, \E Z_1^4<\infty$, then
\begin{displaymath}
 \diccThe(X_3,Y_3)
 \ \pto \
\left(
1+\alpha\left(\frac{\E X_1^2}{\E X_1}+Y_3\right)
\frac{(\E Z_1^3)(\E Z_1^2)}{\E Z_1^4}
+
\alpha^2
Y_3
\frac{\E X_1^2}{\E X_1}
\frac{(\E Z_1^2)^3}{\E Z_1^4}
\right)^{-1}.
\end{displaymath}

\item \label{ite:TLDC2} If $\E Z_1^4<\infty$, then
\begin{displaymath}
 \diccThe(X,Y)
 \ \pto \
\left(
1
+
\alpha
(X_1+Y_3)
\frac{(\E Z_1^3)(\E Z_1^2)}{\E Z_1^4}
+
\alpha^2
X_1Y_3
\frac{(\E Z_1^2)^3}{\E Z_1^4}
\right)^{-1}.
\end{displaymath}
\end{enumerate}
\end{theorem}

Note that for large $Y_3$, the clustering coefficient $\diccThe(X_3,Y_3) = \diccThe(Y_3)$ 
scales as $Y_3^{-1}$.
Similarly, for large $X_1$ and $Y_3$, the probability 
$\diccThe(X,Y)$ scales as $X_1^{-1}Y_3^{-1}$. We remark that 
similar scaling of a related clustering coefficient in an undirected random intersection graph has been observed
in \cite{Deijfen_Kets_2009}.

\begin{remark}
When all attribute weights are equal to a constant $z > 0$, the statement in Theorem~\ref{TLDC}:\eqref{ite:TLDC2}
simplifies into $\diccThe(X,Y) \pto \bigl(1+\alpha z X_1\bigr)^{-1}\bigl(1+\alpha z Y_3\bigr)^{-1}$, a result reported in \cite{Leskela_2015_Istanbul}.
\end{remark}

\begin{remark}
Theorems \ref{T1}, \ref{TDC}, and \ref{TLDC} do not impose any restrictions on the correlation structure of the supply and demand indicators defined by (\ref{rrr}).
\end{remark}

\subsection{Diclique versus transitivity clustering}

An interesting question is to compare the diclique clustering coefficient $\diccThe$ with the commonly used  transitive closure clustering coefficient
\begin{displaymath}
 \trccThe = \PP\bigl(2 \to 4 \, \bigl| \, 2 \to 3 \to 4\bigr),
\end{displaymath} 
see e.g.~\cite{Fagiolo_2007,Wasserman_Faust_1994}.  The next result illustrates  that $\trccThe$ depends heavily on the correlation between the supply and demand indicators characterized by the function $r(x,y,z,\gamma)$ in (\ref{rrr}). A similar finding for a related random intersection graph has been discussed in~\cite{Bloznelis_2010_Random}. We denote $\min\{a,b\}=a\wedge b$.

\begin{theorem}
\label{TC} Let $m,n\to \infty$.
Assume that $\gamma = (nm)^{-1/2}$ and $m/n\to \beta$ for some $\beta > 0$.
Suppose also that $\E X_1^2, \E Y_1^2, \E Z_1^2<\infty$.
\begin{enumerate}[(i)]
\item \label{ite:TC1} If $r(x,y,z,\gamma) = (\gamma x z \wedge 1)(\gamma y z \wedge 1)$,
then $\trccThe \to 0$. 
\item \label{ite:TC2} If $r(x,y,z,\gamma) = \epsilon (\gamma x z \wedge \gamma y z \wedge 1)$ for some $0 < \varepsilon \le 1$
and $\E (X_1\wedge Y_1)>0$, then
\begin{equation}
 \trccThe
 \ \to \ \left( 
1+
\frac{\sqrt{\beta}}{\varepsilon} \frac{\E(X_1 Y_1)}{\E(X_1 \wedge Y_1)} 
\frac{(\E Z_1^2)^2}{\E Z_1^3} 
\right)^{-1}.
\end{equation}
\end{enumerate}
\end{theorem}

The assumption in \eqref{ite:TC1} means that the supply and demand indicators of any particular 
node--attribute pair are conditionally independent given the weights. 
In contrast, the assumption in \eqref{ite:TC2} forces a strong correlation 
between the supply and demand indicators. We note that
condition (\ref{rrr+}) is satisfied in case \eqref{ite:TC2} for all $i \le n$ and $k \le m$ with high probability as $n,m\to \infty$, because $n^{-1/2} \max_{i \le n}(X_i+Y_i) \pto 0$ and $m^{-1/2} \max_{k \le m} Z_k \pto 0$ imply that $\gamma X_i Z_k + \gamma Y_i Z_k \le 1$ for all $i \le n$ and $k \le m$ with high probability.

We remark that in case \eqref{ite:TC1}, and in case \eqref{ite:TC2} with a very small $\varepsilon$, the 
transitive closure clustering coefficient $\trccThe$ becomes negligibly small, 
whereas the diclique clustering coefficient $\diccThe$ 
remains bounded away from zero. 
Hence, it make sense to consider the event 
$\{1 \to 3, \, 1 \to 4, \, 2 \to 3\}$ as a more robust 
predictor of the link $2 \to 4$  than 
the event $\{2 \to 3 \to 4\}$. 
This conclusion has been empirically confirmed for various 
real-world networks in 
\cite{Milo_Shen-Orr_Itzkovitz_Kashtan_Chklovskii_Alon_2002,Zhang_Lu_Wang_Zhu_Zhou_2013}.

\section{Proofs}

The proof of Theorem \ref{T1} goes along similar lines as that of Theorem 1 in~\cite{Bloznelis_Damarackas_2013}. It is
  omitted.  We only give the proofs of Theorems \ref{TDC} and  \ref{TLDC}.
The proof of Theorem \ref{TC} is given in an extended version of the 
paper available from the authors.

We assume for notational convenience
that $\gamma=\alpha m^{-1}$.
Denote events ${\cal A}=\{1\to 3,1\to 4, 2\to 3\}$, ${\cal B}=\{2\to 4\}$  and random variables 
\begin{eqnarray}
\nonumber
&&
 \tilde p_{ik}=\alpha\frac{X_iZ_k}{m},
 \qquad
 \tilde q_{ik}=\alpha\frac{Y_iZ_k}{m}.
\end{eqnarray}
By ${\tilde \PP}$ and ${\tilde \E}$ we denote the conditional probability and  expectation given $X,Y,Z$. 
Note that $p_{ik} = {\tilde \PP}({\mathbb I}_{i\to k}=1)$, $q_{ik} ={\tilde \PP}({\bf I}_{k\to i}=1)$, and
\begin{equation}\label{pq2}
p_{ik}
 =
1\wedge \tilde p_{ik},
\qquad
q_{ik}
=
1\wedge \tilde q_{ik}.
\end{equation}

\begin{proof}[of Theorem \ref{TDC}]
We observe that ${\cal A}=\cup_{i\in[4]}{\cal A}_i$, where
\begin{eqnarray}\nonumber
&&
{\cal A}_1=
\quad
\
\,
\bigcup_{k\in {\cal C}_1}{\cal A}_{1.k},
\qquad
\quad
\quad
\,
{\cal A}_{1.k}
=
\bigl\{
{\mathbb I}_{1\to k}{\mathbb I}_{2\to k}{\bf I}_{k\to 3}
{\bf I}_{k\to 4}
=1
\bigr\}, 
\\
\nonumber
&&
{\cal A}_2=
\
\
\bigcup_{(k,l)\in {\cal C}_2}{\cal A}_{2.kl},
\qquad
\quad
\
{\cal A}_{2.kl}
=
\bigl\{
{\mathbb I}_{1\to k}{\mathbb I}_{2\to l}{\bf I}_{k\to 3}
{\bf I}_{k\to 4}{\bf I}_{l\to 3}
=1
\bigr\}, 
\\
\nonumber
&&
{\cal A}_3=
\
\
\bigcup_{(k,l)\in {\cal C}_3}{\cal A}_{3.kl},
\qquad
\quad
\
{\cal A}_{3.kl}
=
\bigl\{
{\mathbb I}_{1\to k}{\mathbb I}_{1\to l}{\mathbb I}_{2\to k}
{\bf I}_{k\to 3}{\bf I}_{l\to 4}
=1
\bigr\}, 
\\
\nonumber
&&
{\cal A}_4=\bigcup_{(j,k,l)\in {\cal C}_4}{\cal A}_{4.jkl},
\qquad
\quad
{\cal A}_{4.jkl}
=
\bigl\{
{\mathbb I}_{1\to j}{\mathbb I}_{1\to k}{\mathbb I}_{2\to l}
{\bf I}_{j\to 3}{\bf I}_{k\to 4} {\bf I}_{l\to 3}
=1
\bigr\}. 
\end{eqnarray}
Here
${\cal C}_1=[m]$,
$
{\cal C}_2={\cal C}_3=\{(k,l):\,k\not=l;\, k,l\in [m]\}$, and
$ 
{\cal C}_4=\{(j,k,l):\, j\not=k\not=l;\, j,k,l\in [m]\}$.
Hence, by inclusion--exclusion,
\begin{displaymath}
\sum_{i\in[4]}\PP({\cal A}_i)
-
\sum_{\{i,j\}\subset [4]}\PP({\cal A}_i\cap{\cal A}_j)
\le
 \PP({\cal A})
\le 
\sum_{i\in[4]}\PP({\cal A}_i).
\end{displaymath}
We prove the theorem in Claims $1-3$ below. Claim $2$ implies that 
$\PP({\cal A})= \sum_{i\in [4]}\PP({\cal A}_i)+O(m^{-4})$.
Claim $3$ implies that 
$\PP({\cal A}\cap{\cal B})=\PP({\cal A}_1)+O(m^{-4})$. Finally, Claim $1$
establishes the approximation (\ref{DC}) to the ratio
$\diccEmp=\PP({\cal A}\cap{\cal B})/\PP({\cal A})$.

{\it Claim 1. We have }
\begin{eqnarray}\label{A1-3}
&&
\PP({\cal A}_1)=\alpha^4m^{-3}A_1(1+o(1)),
\\
\label{A2-1}
&&
\PP({\cal A}_2)
=
\alpha^5m^{-3}A_2 (1+o(1)),
\\
\label{A3-1}
&&
\PP({\cal A}_3)
=
\alpha^5m^{-3}A_3 (1+o(1)),
\\ 
\label{A4-1}
&&
\PP({\cal A}_4)
=
 \alpha^6m^{-3}A_4(1+o(1)).
\end{eqnarray}
Here we denote
\begin{displaymath}
A_1=a_1^2b_1^2h_4,
\quad
A_2=a_1^2b_1b_2h_2h_3,
\quad
A_3=a_1a_2b_1^2h_2h_3,
\quad
A_4=a_1a_2b_1b_2h_2^3.
\end{displaymath}
and $a_r=\E X_1^r$, $b_4=\E Y_1^r$, $h_r=\E Z_1^r$.

{\it Claim 2.} For $1\le i<j\le 4$ we have
\begin{equation}
 \PP({\cal A}_i\cap{\cal A}_j)=O(m^{-4}).
\end{equation}

{\it Claim 3.}  We have
\begin{equation}\label{B}
 \PP({\cal B}\cap{\cal A})=\PP({\cal A}_1)+O(m^{-4}).
\end{equation}

{\it Proof of Claim 1.} 
We estimate  every $\PP({\cal A}_r)$ using inclusion-exclusion $I_1-I_2\le \PP({\cal A}_r)\le I_1$.
Here
\begin{displaymath}
I_1=I_1(r)=\sum_{x\in {\cal C}_r}\PP({\cal A}_{r.x}),
\qquad
I_2
=
I_2(r)=
\sum_{\{x,y\}\subset {\cal C}_r}
\PP({\cal A}_{r.x}\cap{\cal A}_{r.y}).
\end{displaymath}
 Now
 (\ref{A1-3}-\ref{A4-1}) follow from the  approximations
\begin{eqnarray}\label{A8}
&&
I_1=\alpha^4m^{-3}A_1(1+o(1)),
\qquad
\qquad
\,
I_2
=
\alpha^5m^{-3}A_2 (1+o(1)),
\\
\nonumber
&&
I_3
=
\alpha^5m^{-3}A_3 (1+o(1)),
\qquad
\quad
\
I_4
=
\alpha^6m^{-3}A_4(1+o(1))
\end{eqnarray} 
and  bounds $I_2(r)=o(m^{-3})$, for $1\le r\le 4$.

Firstly we show (\ref{A8}). We only prove the first relation. The remaining cases are treated in much the same way. 
From the inequalities, see (\ref{pq2}),
\begin{eqnarray}\nonumber
&&
\tilde p_{1k}\tilde p_{2k}\tilde q_{3k}\tilde q_{4k}
\ge
 p_{1k}p_{2k}q_{3k}q_{4k}
\ge
\tilde p_{1k}\tilde p_{2k}\tilde q_{3k}\tilde q_{4k}{\mathbb I}_k'
\ge 
\tilde p_{1k}\tilde p_{2k}\tilde q_{3k}\tilde q_{4k}
-
\tilde p_{1k}\tilde p_{2k}\tilde q_{3k}\tilde q_{4k}{\mathbb I}_k^*,
\\
\nonumber
&&
{\mathbb I}_k'
=
{\mathbb I}_{\tilde p_{1k}\le 1}
{\mathbb I}_{\tilde p_{2k}\le 1}
{\mathbb I}_{\tilde q_{3k}\le 1}
{\mathbb I}_{\tilde q_{4k}\le 1},
\qquad
{\mathbb I}_k^*
=
{\mathbb I}_{\tilde p_{1k}> 1}
+
{\mathbb I}_{\tilde p_{2k}> 1}
+
{\mathbb I}_{\tilde q_{3k}> 1}
+
{\mathbb I}_{\tilde q_{4k}> 1},
\end{eqnarray}
 we obtain that
\begin{equation}\label{A111}
\PP({\cal A}_{1.k})
=
\E p_{1k}p_{2k}q_{3k}q_{4k}
=
\E \tilde p_{1k}\tilde p_{2k}\tilde q_{3k}\tilde q_{4k}
+R,
\end{equation}
where
\begin{displaymath}
\E \tilde p_{1k}\tilde p_{2k}\tilde q_{3k}\tilde q_{4k}
=\alpha^4m^{-4}A_1
\qquad
{\text{and}}
\qquad
|R|
\le  
\E \tilde p_{1k}\tilde p_{2k}\tilde q_{3k}\tilde q_{4k}{\mathbb I}_k^*
=o(m^{-4}).
\end{displaymath}
Hence $I_1=m\PP({\cal A}_{1.k})=\alpha^4m^{-3}A_1(1+o(1))$. 

\medskip

Secondly we show that $I_2(r)=o(m^{-3})$, for $1\le r\le 4$. For $r=1$ 
the bound $I_2(1)
=
{\binom{m}{2}}\PP({\cal A}_{1.k}\cap {\cal A}_{1.l})=o(m^{-3})$
follows from
the inequalities 
\begin{displaymath}
\PP({\cal A}_{1.k}\cap {\cal A}_{1.l})
\le
\E
 \tilde p_{1k}\tilde p_{2k}\tilde q_{3k}\tilde q_{4k}
 \tilde p_{1l}\tilde p_{2l}\tilde q_{3l}\tilde q_{4l}
 =O(m^{-8}).
\end{displaymath}
For $r=2,3$ we split $I_2(r)=J_1+\dots+J_5$, 
where
\begin{eqnarray}\nonumber
&&
J_1
=
\sum_{\{(k,l),(k,l')\}\subset {\cal C}_r}
\PP({\cal A}_{r.kl}\cap {\cal A}_{r.kl'}),
\qquad
\,
J_2
=
\sum_{\{(k,l),(k',l)\}\subset {\cal C}_r}
\PP({\cal A}_{r.kl}\cap {\cal A}_{r.k'l}),
\\
\nonumber
&&
J_3
=
\sum_{\{(k,l),(k',l')\}\subset {\cal C}_r}
\PP({\cal A}_{r.kl}\cap {\cal A}_{r.k'l'}),
\qquad
J_4
=
\sum_{\{(k,l),(k',k)\}\subset {\cal C}_r,\, k'\not=l}
\PP({\cal A}_{r.kl}\cap {\cal A}_{r.k'k}),
\\
\nonumber
&&
J_5
=
\sum_{(k,l)\in {\cal C}_r}
\PP({\cal A}_{r.kl}\cap {\cal A}_{r.lk}).
\end{eqnarray}
In the first (second) sum distinct pairs $x=(k,l)$ and $y=(k',l')$ share the 
first (second) coordinate. In the third sum all coordinates of the pairs
$(k,l),(k',l')$ are different. 
In the fourth sum the pairs $(k,l),(k',k)$ only
share one common element, but it appears in different coordinates.  
We show that each  $J_i=o(m^{-3})$. Next we only consider the case of $r=2$. 
The case  of  $r=3$ is treated in a similar way. We have
\begin{eqnarray}\nonumber
&&
J_1
=
m{\binom{m-1}{2}}\PP({\cal A}_{2.kl}\cap{\cal A}_{2.kl'})
\le 
m^3 \E H_1,
\quad
 H_1=p_{1k}p_{2l}p_{2l'}q_{3k}q_{4k}q_{3l}q_{3l'},
\\
\nonumber
&&
J_2
=
m{\binom{m-1}{2}}\PP({\cal A}_{2.kl}\cap{\cal A}_{2.k'l})
\le 
m^3 \E H_2,
\quad H_2=p_{1k}p_{1k'}p_{2l}q_{3k}q_{4k}q_{3k'} q_{4k'}q_{3l},
\\
\nonumber
&&
J_3
=
\binom{m}{2}{\binom{m-2}{2}}\PP({\cal A}_{2.kl}\cap{\cal A}_{2.k'l'})
\le 
m^4 \E H_3,
\quad
H_3= p_{1k}p_{1k'}p_{2l}p_{2l'}q_{3k}q_{4k}q_{3k'}q_{4k'}q_{3l}q_{3l'},
\\
\nonumber
&&
J_4
=
m(m-1)(m-2)\PP({\cal A}_{2.kl}\cap{\cal A}_{2.k'k})
\le 
m^3 \E H_4,
\quad
H_4= p_{1k}p_{1k'}p_{2l}p_{2k}q_{3k}q_{4k}q_{3k'}q_{4k'}q_{3l},
\\
\nonumber
&&
J_5
=
\binom{m}{2}\PP({\cal A}_{2.kl}\cap{\cal A}_{2.lk})
\le 
m^2 \E H_5,
\quad
H_5= p_{1k}p_{1l}p_{2l}p_{2k}q_{3k}q_{3l}q_{4k}q_{4l}.
\end{eqnarray}

In the product $H_1$ we estimate the typical  factors $p_{ij}\le \tilde p_{ij}$ and $q_{ij}\le \tilde q_{ij}$, but  
\begin{equation}\label{trunc}
q_{3l}
\le 
\tilde q_{3l}{\mathbb I}_{Y_3\le\sqrt{m}}
+
{\mathbb I}_{Y_3> \sqrt{m}}
\le 
\alpha m^{-1/2}Z_l
+
{\mathbb I}_{Y_3>\sqrt{m}}.
\end{equation}
We obtain
\begin{equation}\label{H3}
\E H_1
\le 
\alpha^6
m^{-6}a_1a_2b_1h_2h_3
\bigl(
b_2h_2\alpha m^{-1/2}+h_1\E Y_3^2{\mathbb I}_{Y_3>\sqrt{m}}\bigr)
=
o(m^{-6}).
\end{equation}
 Hence $J_1=o(m^{-3})$.
Similarly, we show that $J_2=o(m^{-4})$. 
Furthermore, while estimating $H_3$ we apply (\ref{trunc}) 
to $q_{3l}$ and $q_{3l'}$
and apply $p_{ij}\le \tilde p_{ij}$ and $q_{ij}\le \tilde q_{ij}$
to remaining factors.
We obtain 
\begin{equation}\label{H3+}
H_3
\le 
\tilde p_{1k}\tilde p_{1k'}\tilde p_{2l}\tilde p_{2l'}\tilde q_{3k}
\tilde q_{4k}\tilde q_{3k'}\tilde q_{4k'}
(\alpha m^{-1/2}Z_l
+
{\mathbb I}_{Y_3>\sqrt{m}})
(\alpha m^{-1/2}Z_{l'}
+
{\mathbb I}_{Y_3>\sqrt{m}}).
\end{equation} 
Since the expected value of the product on the right is 
$o(m^{-8})$, we conclude that  $\E H_3=o(m^{-8})$. Hence $J_3=o(m^{-4})$.
Proceeding in a similar way we establish the bounds 
$J_4=o(m^{-5})$ and $J_5=O(m^{-6})$. 

We explain the truncation step (\ref{trunc}) in some more detail. 
A simple upper bound for $H_1$ is the product
\begin{displaymath}
\tilde p_{1k}\tilde p_{2l}\tilde p_{2l'}
\tilde q_{3k}\tilde q_{4k}\tilde q_{3l}\tilde q_{3l'}
=
\alpha^7m^{-7}X_1X_2^2Y_3^3Y_4Z_k^3Z_l^2Z_{l'}^2.
\end{displaymath}
It contains an undesirable high power $Y_3^3$. Using (\ref{trunc}) 
instead of the 
simple upper bound $q_{3l}\le\tilde q_{3l}$ we have reduced in (\ref{H3}) 
the power of $Y_3$ down to  $2$.  Similarly, in (\ref{H3+}) we have reduced 
the power of $Y_3$ from $4$  to  $2$.
            
Using the truncation argument we obtain the upper bound
$I_2(4)=o(m^{-3})$ under moment conditions $\E X_1^3, \E Y_1^3, \E Z_1^4<\infty$.
The proof is similar to that of the bound  $I_2(2)=o(m^{-3})$ above. 
We omit routine, but tedious calculation. 

\medskip

{\it Proof of Claim 2.}
We only prove that $q:=\PP({\cal A}_3\cap{\cal A}_4)=O(m^{-4})$. 
The remaining cases are 
treated in a similar way. 
For $x=(j,k,l)\in {\cal C}_4$ and $y=(r,t)\in {\cal C}_3$
we denote, for short, 
$
{\mathbb I}_{{\cal A}_{4.x}}={\mathbb I}^*_{x}={\mathbb I}^*_{jkl}$ and
$
{\mathbb I}_{{\cal A}_{3.y}}={\mathbb I}_{y}={\mathbb I}_{rt}$.
For $q=\E {\mathbb I}_{{\cal A}_4}{\mathbb I}_{{\cal A}_3}$, we write, 
by the symmetry,
\begin{displaymath}
q
\le
\E \Bigl(\sum_{x\in{\cal C}_4}{\mathbb I}^*_x\Bigr)
{\mathbb I}_{{\cal A}_3}
=
m(m-1)(m-2)
\E {\mathbb I}^*_{123}
{\mathbb I}_{{\cal A}_3}
\end{displaymath}
and 
\begin{displaymath}
\E {\mathbb I}^*_{123}
{\mathbb I}_{{\cal A}_3}
\le 
\E {\mathbb I}^*_{123}\Bigl(\sum_{y\in{\cal C}_3}{\mathbb  I}_y\Bigr)
=
\E {\mathbb I}^*_{123}(J_1+J_2+J_3).
\end{displaymath}
Here
\begin{displaymath}
J_1
=
\sum_{r,t\in[m]\setminus[3],\, r\not=t}
{\mathbb I}_{rt},
\qquad
J_2
=
\sum_{r\in[m]\setminus[3]}
\
\
\sum_{s\in [3]}
\bigl(
{\mathbb I}_{sr}+{\mathbb I}_{rs}
\bigr),
\qquad
J_3=\sum_{r,t\in [3],\, r\not=t} {\mathbb I}_{rt}.
\end{displaymath} 
Finally, we  show that 
$\E {\mathbb I}^*_{123}J_i=O(m^{-7})$, $i\in [3]$.
For $i=1$ we have, by the symmetry,
\begin{eqnarray}
 \E {\mathbb I}^*_{123}J_1=(m-3)(m-4)\E {\mathbb I}^*_{123}{\mathbb I}_{45}.
\end{eqnarray}
Invoking the inequalities
\begin{equation}\label{truncation+1}
\E {\mathbb I}^*_{123}{\mathbb I}_{45}
=
\E {\tilde \E}{\mathbb I}^*_{123}{\mathbb I}_{45}
\le 
\E
\tilde p_{11}\tilde p_{12}
\tilde p_{15}
\tilde p_{23}\tilde p_{24}
\tilde q_{13}\tilde q_{24}\tilde q_{33}\tilde q_{43}\tilde q_{54}
=O(m^{-10}) 
\end{equation}
we obtain
$\E {\mathbb I}^*_{123}J_1=O(m^{-8})$.

The bound $\E {\mathbb I}^*_{123}J_2=O(m^{-7})$ is obtained 
from the identity (which follows by symmetry)
\begin{displaymath}
\E {\mathbb I}^*_{123}J_2
=
(m-3)
\sum_{s\in [3]}
\bigl(
\E {\mathbb I}^*_{123} {\mathbb I}_{s4}+\E {\mathbb I}^*_{123}{\mathbb I}_{4s}
\bigr),
\end{displaymath}
combined with  bounds
$\E {\mathbb I}^*_{123} {\mathbb I}_{s4}
+
\E {\mathbb I}^*_{123}{\mathbb I}_{4s}
=O(m^{-8})$,  $s\in[3]$. 
We only show the latter bound for $s=3$. 
The cases $s=1,2$ are treated in a similar way.
 We have
\begin{eqnarray}
\nonumber
&&
\E {\mathbb I}^*_{123} {\mathbb I}_{34}
\le
\E \tilde p_{11}\tilde p_{12}\tilde p_{13}\tilde p_{23}
\tilde q_{13}\tilde q_{24}\tilde q_{33}\tilde q_{44}
=O(m^{-8}),
\\
\nonumber
&&
\E {\mathbb I}^*_{123} {\mathbb I}_{43}
\le
\E \tilde p_{11}\tilde p_{12}\tilde p_{13}\tilde p_{23}\tilde p_{24}
\tilde q_{13}\tilde q_{24}\tilde q_{33}\tilde q_{34}\tilde q_{43}=O(m^{-10}).
\end{eqnarray}
The proof of $\E {\mathbb I}^*_{123}J_3=O(m^{-7})$ is similar. 
It is omitted. 

\medskip

{\it Proof of Claim 3.} 
We use the notation 
${\overline{\mathbb I}}_{{\cal A}_j}=1-{\mathbb I}_{{\cal A}_j}$ for the indicator 
of the event ${\overline{\cal A}_j}$ complement to ${\cal A}_j$.  
For $2\le i\le 4$ we denote ${\cal H}_i
= 
\bigl({\cal A}_i\cap{\cal B}\bigr)\setminus \cup_{1\le j\le i-1}{\cal A}_j$.
 We have
\begin{displaymath}
 \PP({\cal A}\cap{\cal B})=\PP(\cup_{i\in[4]}{\cal A}_i\cap {\cal B})
=
\PP({\cal A}_1\cap{\cal B})+R,
\qquad
0\le R\le \PP(\cup_{2\le i\le 4}{\cal H}_i).
\end{displaymath}
Note that $\PP({\cal A}_1\cap{\cal B})=\PP({\cal A}_1)$. It remains to show 
that $\PP({\cal H}_i)=O(m^{-4})$, 
$2\le i\le 4$.

We have, by the symmetry,
\begin{eqnarray}\label{H2++}
 \PP({\cal H}_2)
=
\E 
{\mathbb I}_{{\cal A}_2}
{\mathbb I}_{{\cal B}}
{\overline {\mathbb I}}_{{\cal A}_1}
\le
\E
\sum_{x\in{\cal C}_2}{\mathbb I}_{{\cal A}_{2.x}} 
{\mathbb I}_{{\cal B}}
{\overline {\mathbb I}}_{{\cal A}_1}
=
m(m-1)
\E {\mathbb I}_{{\cal A}_{2.12}} 
{\mathbb I}_{{\cal B}}
{\overline {\mathbb I}}_{{\cal A}_1}.
\end{eqnarray}
Furthermore, we have ${\mathbb I}_{{\cal A}_{2.12}} 
{\mathbb I}_{{\cal B}}
{\overline {\mathbb I}}_{{\cal A}_1}
\le
{\mathbb I}_{{\cal A}_{2.12}} 
\bigl(
{\bf I}_{2\to 4}
+
\sum_{3\le j\le m}
{\mathbb I}_{2\to j}{\bf I}_{j\to 4}
\bigr)$
and, by the symmetry,
\begin{displaymath}
\E {\mathbb I}_{{\cal A}_{2.12}} 
{\mathbb I}_{{\cal B}}
{\overline {\mathbb I}}_{{\cal A}_1}
\le
\E  {\mathbb I}_{{\cal A}_{2.12}} {\bf I}_{2\to 4}
+
(m-2)\E  {\mathbb I}_{{\cal A}_{2.12}}{\mathbb I}_{2\to 3}{\bf I}_{3\to 4}.
\end{displaymath}
A simple calculation shows that 
$\E {\mathbb I}_{{\cal A}_{2.12}} {\bf I}_{2\to 4}
\le 
\E \tilde p_{11}\tilde p_{22}\tilde q_{13}\tilde q_{14}\tilde q_{23}\tilde q_{24}
=
O(m^{-6})$. Similarly, 
$\E  {\mathbb I}_{{\cal A}_{2.12}}{\mathbb I}_{2\to3}{\bf I}_{3\to 4}=O(m^{-7})$.
Therefore, $\E {\mathbb I}_{{\cal A}_{2.12}}{\mathbb I}_{{\cal B}}
{\overline {\mathbb I}}_{{\cal A}_1} =O(m^{-6})$.
Now (\ref{H2++}) implies
$\PP({\cal H}_2)=O(m^{-4})$. 
The bounds 
$\PP({\cal H}_j)=O(m^{-4})$, $j=3,4$ are  obtained in a similar way.
\end{proof}

\begin{proof}[of  Theorem \ref{TLDC}]
The proof is the same as that of Theorem \ref{TDC}, but while evaluating the 
probabilities of events ${\cal A}$ and ${\cal A}\cap {\cal B}$ we treat
$X_1,X_2,Y_3,Y_4$, respectively $Y_3$, as constants. 
\end{proof}

\bibliographystyle{splncs03}
\bibliography{lslReferences}
\end{document}